\documentclass[polymers,article,submit,moreauthors,pdftex]{absmdpinoline} 

\firstpage{1} 
\pubvolume{1}
\issuenum{1}
\articlenumber{0}
\pubyear{2021}
\copyrightyear{2020}
\datereceived{} 
\dateaccepted{} 
\datepublished{} 
\hreflink{https://doi.org/} 

\Title{Atomic scale mechanisms controlling the oxidation of polyethylene: a first principles study}

\TitleCitation{Atomic scale mechanisms controlling the oxidation of polyethylene: a first principles study}

\Author{Yunho Ahn $^{1}$\orcidA{}, Xavier Colin $^{2}$ and Guido Roma $^{1,}$*}

\AuthorNames{Yunho Ahn, Xavier Colin and Guido Roma}

\AuthorCitation{Ahn, Y.; Colin, X.; Roma, G.}

\address[SRMP]{$^{1}$ \quad Universit\'e Paris-Saclay, CEA, Service de Recherches de M\'etallurgie Physique, 91191 Gif sur Yvette, France}
\address[ENSAM] {$^{2}$ \quad PIMM, Arts et Metiers Institute of Technology, CNRS, CNAM, HESAM University, 151 Boulevard de L'Hopital, 75013, Paris, France}
\address{%
$^{1}$ \quad Universit\'e Paris-Saclay, CEA, Service de Recherches de M\'etallurgie Physique, 91191 Gif sur Yvette, France; yunho.ahn@cea.fr (Y.A.); guido.roma@cea.fr (G.R.)\\
$^{2}$ \quad PIMM, Arts et Metiers Institute of Technology, CNRS, CNAM, HESAM University, 151 Boulevard de L'Hopital, 75013, Paris, France; xavier.colin@ensam.eu (X.C.)}

\corres{Correspondence: guido.roma@cea.fr (G.R.)}

\abstract{  Understanding the degradation mechanisms of aliphatic polymers by thermal oxidation and radio-oxidation is very important in order to assess their lifetime in a variety of industrial applications. We focus here on polyethylene as a prototypical aliphatic polymer. Kinetic models describing the time evolution of the concentration of chain defects and radicals species in the material identify a relevant step in the formation and subsequent decomposition of transient hydroperoxides species, finally leading to carbonyl defects, in particular ketones.
  In this paper we first summarize the most relevant mechanistic paths proposed in the literature for hydroperoxide formation and decomposition and, second, revisit them using first principles calculations based on Density Functional Theory (DFT).
  Our results partially confirm commonly accepted reaction energies, but also propose alternative, more favourable, reaction paths.
  We highlight the influence of the environment ---crystalline or not--- on the outcome of some of the studied chemical reactions.
 A remarkable result of our calculations is that hydroxyl radicals play an important role in the decomposition of hydroperoxides. Based on our findings, it should be possible to improve the set of equations and parameters used in current kinetic simulations of polyethylene radio-oxidation.
}


\begin{document}

\section{Introduction}
\label{Intro}

Polyethylene (PE) is one of the most common polymers and is employed in a
variety of applications such as common packaging plastic, domestic sectors,
automobile, biomedicine, and electric cables insulation, including power
cables in nuclear power plants (NPP)\cite{bombelli_polyethylene_2017,peacock_handbook_2000,fang_processing_2005,plesa_polyethylene_2019}. During
normal operation degradation takes place through internal or external
processes which end up limiting the lifetime of the material. A crucial aging
mechanism of polyethylene, as well as of many polymers, is oxidation; this
process can be initiated by electronic excitations produced by 
irradiation (UV, $\gamma$-rays, electrons, swift heavy
ions)\cite{celina_review_2013,allen_fundamentals_1992,martinez-vega_dielectric_2010}. However, due to the complexity of the
oxidation mechanisms, many studies rely on the assumptions usually made for
kinetic modelling of PE degradation \cite{bolland_kinetic_1946,tobolsky_low_1950,decker_aging_1973,somersall_computer_1985,gillen_kinetic_1985,gillen_general_1995,seguchi_degradation_2011,khelidj_oxidation_2006}. For
example Niki {\it et al.}\cite{niki_aging_1973}, in the first of a series of
three papers, verified the fraction of the oxidation products by thermal
decomposition based on their kinetic models in bulk atactic
polypropylene. Later, other works were devoted to the quantification of the
products under thermo- and photo-oxidation conditions by means of IR
spectroscopy~\cite{costa_ultra_1997,salvalaggio_multi-component_2006,fodor_determination_1984,da_cruz_thermo-oxidative_2016}. In particular, carbonyl groups
(namely ketones) were found to be the majority products under thermo-oxidation
conditions. Despite the fact that ketones can undergo Norrish-type reactions
under photochemical conditions \cite{gardette_photo-_2013}, they still are majority
products \cite{costa_ultra_1997}. The explanation of the kinetic path towards the
formation of ketones and other carbonyl species have sparked a number of
studies based on rate theory models essentially assuming homogeneous
concentration (i.e., homogenous chemical kinetics)\cite{khelidj_oxidation_2006,khelidj_oxidation_2006-1,colin_lifetime_2007,colin_kinetic_2010,colin_about_2003,colin_determination_2004,colin_classical_2006}; in these studies,
important intermediate species include hydroperoxides (POOH groups).
In a recent kinetic study by Da Cruz {\it et al.}\cite{da_cruz_thermo-oxidative_2016}, a new
formation mechanism of ketones starting from two POOH is suggested when the
bimolecular decomposition of POOH is the main source of radicals.

In the last decade, the idea of investigating the elementary mechanisms of PE
degradation using {\sl ab initio} calculations has emerged. Although kinetic
constants and activation energies have been generally deduced from experiments
by indirectly measuring the concentration of the products and assuming
Arrhenius behaviour, there are reactions and processes which are difficult to
probe experimentally such as H diffusion, abstraction, or parameters like
transition ring size; the corresponding (free) energy barriers can be predicted by first principles
calculations\cite{hayes_kinetic_2009,chan_self-abstraction_1998,kysel_dft_2011,oluwoye_oxidation_2015,pfaendtner_quantum_2006,de_sainte_claire_degradation_2009}. Especially,
calculations have focused on the activation energy of reactions, because it
crucially controls kinetic rate constants. However, describing PE under
realistic conditions from first principles is a difficult endeavour, because of a complex
microstructure which cannot be faithfully described by models whose size is
limited to few tenths of atoms. Most studies focused on single molecular units
in gas phase and the search of associated transition states; other atomic
environments such as crystalline/amorphous lamell\ae\ of polyolefins and their
interfaces have not, or rarely\cite{ceresoli_trapping_2004,unge_space_2013},
been tackled with {\sl ab initio} methods. In particular, reaction processes in
crystalline PE have been rarely considered, even though carbonyl defects and
other species obviously affect the optical and electrical properties by
forming shallow and deep
traps\cite{ceresoli_trapping_2004,huzayyin_quantum_2010,unge_space_2013,roma_optical_2018}. Besides,
the formation of defects in PE during radio-oxidation occurs due
to the relatively low energy barrier for oxygen permeability, which is of the
order of 0.4 eV\cite{wang_ethylene_1998}, diffusing from the surface all the way through the amorphous
regions; although it is generally assumed that crystalline regions are impermeable to oxygen at room temperature,\cite{Rogers-Comyn} the relationships between crystallinity, density, and oxygen permeation at various conditions are not fully understood\cite{wang_ethylene_1998}. Not only the permeability plays a role, but also the pristine concentration of carbonyl groups does, which in commercial PE applications can reach 0.1\%, resulting from the presence of oxygen impurities during polymerization\cite{huzayyin_quantum_2010}.
It is therefore necessary to advance the atomic scale understanding of
relevant chemical reactions through various models approaching crystalline and
amorphous regions and to compare them to results for gas phase molecules. In
this work, 
guided by the kinetic scheme focusing on the production of ketones through the intermediate hydroperoxide species, we investigate the
activation energies by calculating full reaction paths using the climbing
image nudged elastic band (CI-NEB)\cite{henkelman_climbing_2000} method. 

We discuss three main reaction pathways:
\begin{enumerate}[label=\roman*)]
 \item the capture of oxygen by an alkyl radical,
 \item the formation of hydroperoxides, 
 \item the decomposition of the latter. 
 \end{enumerate}
Two models are considered: small molecules and crystalline PE. Although relevant reactions are supposed to occur mostly in the amorphous region, bimolecular reactions barriers calculated in the crystal can, in many cases, be safely transposed to the amorphous, as we will show. 
Reactions involving a hydroxyl radical are found to be particularly relevant for
the decomposition of hydroperoxides and occur spontaneously regardless of the
position of hydroxyl radical, at variance with reactions where hydroxyls are
not present. Finally, we conclude that this radical is the best candidate
leading to alkyl radical chain oxidation and the degradation of PE thanks to its reactivity.

\section{Computational details}
\label{Method}

 All the results presented in this paper are based on density functional
 theory (DFT). Equilibrium structures and their total energies are obtained
 from the \textsc{Quantum-Espresso} software package by using the PWSCF
 module\cite{giannozzi_quantum_2009}. The energy barriers and reaction pathways are calculated by
 the climbing image nudged elastic band (CI-NEB) method, as implemented in the
 \textsc{Quantum-Espresso} distribution.

 We rely on two models, the first one based on isolated molecules of varying length, the second one is crystalline PE, in the orthorhombic structure, as described in Ref. \cite{roma_optical_2018}.
 We will call them in the following molecular and solid models,
 respectively. While reactions occurring on isolated molecules might well represent unimolecular reactions occurring in low density regions of the amorphous, the crystal is the only one able to describe intermolecular (namely bimolecular) reactions. In some specific cases we will show the influence of density on reaction barriers, to assess the distribution of reaction energies due to local density variations.

\begin{figure}
   \includegraphics[width=0.7\columnwidth]{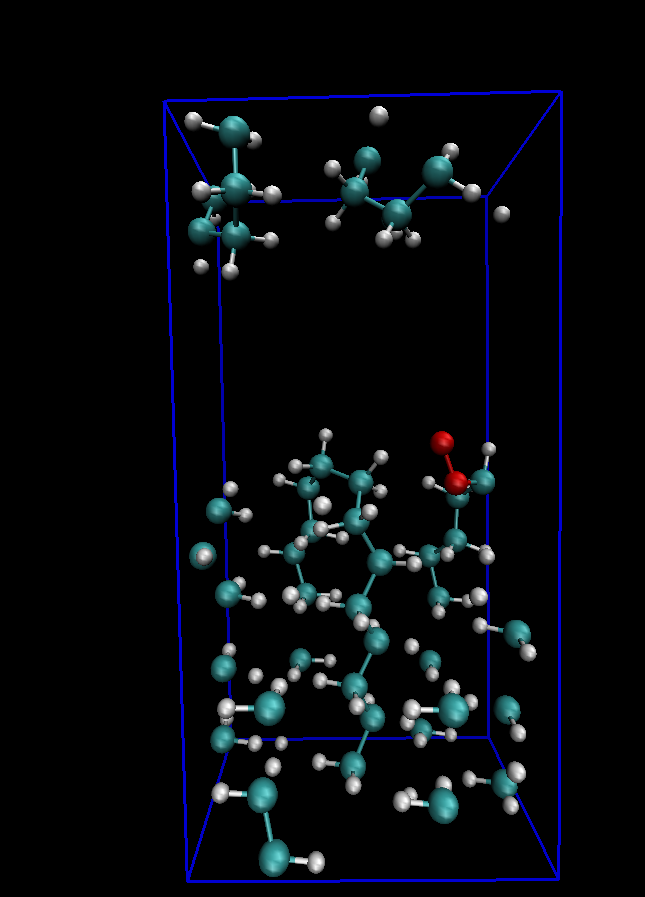}
   \caption{The periodically repeated 133-atom unit cell used to mimick the interface between two crystalline lamellae of PE, here with a peroxy radical.}
   \label{FigLamellar}
   \end{figure}

 For a few reactions we calculated the migration barrier in a different model mimicking the interface between two crystalline regions, with bent polymer chains at the crystal surfaces; the two facing surfaces are separated by approximately 5~\AA. The unit cell of the lamellar model, containing 132 atoms for pure PE, is shown in Figure~\ref{FigLamellar}, similar to a model previously used to describe a carboxyl group grafted onto a lamella\cite{ceresoli_trapping_2004}.

 Pseudopotentials, norm-conserving (nc) for C and H and both nc and
 ultrasoft for oxygen, were generated as described in \cite{roma_optical_2018} and
 used with the optB86b+vdW exchange-correlation (xc) functional\cite{klimes_van_2011} and,
 in some cases, with the hybrid functional vdW-DF-cx0\cite{berland_assessment_2017}.

The functional optB86b+vdW includes a gradient corrected short range xc contribution and a long-range
non-local van der Waals correlation term in order to provide a good description of
van der Waals interchain interactions in crystalline PE. In addition, for the
accuracy of some specific reactions, we used the hybrid functional vdW-DF-cx0
which mixes into the same exchange part, a portion of Hartree-Fock exchange
energy to improve the description of electronic density localization.  
 For molecular models, we used body-centred tetragonal unit cells in order to
 maximise the chain ends distance between periodic images of the molecules;
 the unit cells were 40 $\times$ 40 bohr wide in the plane perpendicular to the chain
 and they exceeded the molecules length in the direction parallel to the chain
 so to have at least a distance of 25 bohr between atoms of two different
 periodic images.  The unit cell of the crystalline solid is orthorhombic,
 containing 12 atoms, and we sampled the BZ with a 3$\times$3$\times$6 $\Gamma$-centred regular
 {\bf k}-point mesh. To model isolated defects in the solid we used a 2$\times$1$\times$4
 supercell (containing 96 atoms plus defects), and we employed a 2$\times$3$\times$2
 $\Gamma$-centred {\bf k}-point mesh. The theoretical equilibrium lattice parameters of the orthorhombic unit cell in atomic units were: a=9.18 Bohr, b=13.14 Bohr, and 4.84 Bohr, and for 2$\times$1$\times$4 supercell, a=18.36 Bohr, b=13.14 Bohr, and c=19.36 Bohr.
 As our NEB calculations revealed a quite low migration barrier for the oxygen molecule in the crystal, we performed a Car-Parrinello molecular dynamics simulation in the NVT ensemble using the {\tt cp.x} executable in the \textsc{Quantum-Espresso} distribution. The simulation run for 4.6~ps with a time step of 0.12~fs and a fictitious electronic mass of 300~atomic units.

\section{Results and discussion}
\label{Results}

\subsection{Oxygen capture by an alkyl radical}
\label{OCapture}

\begin{figure}
\includegraphics[width=\columnwidth]{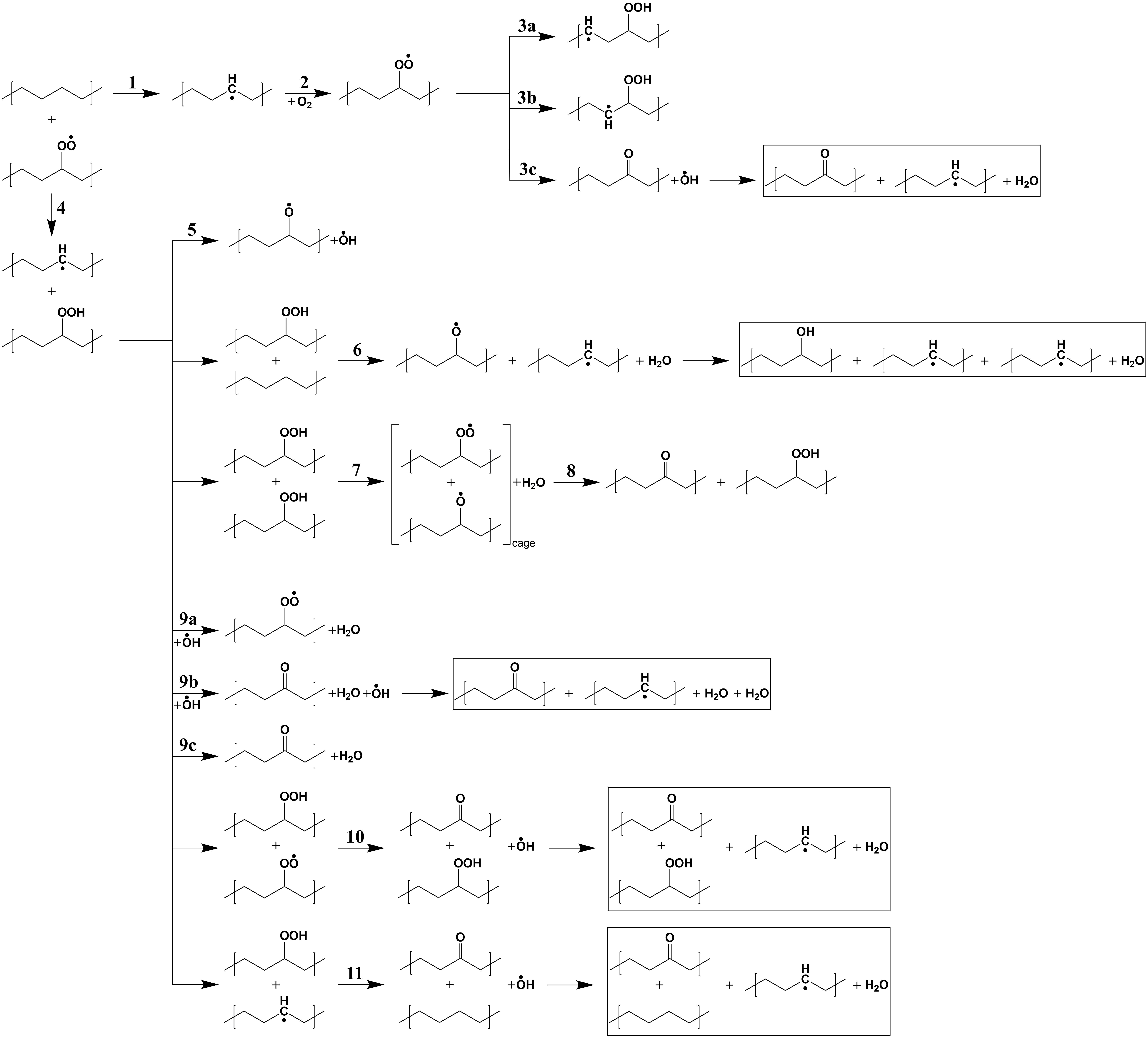}
\caption{A summary of reaction paths leading to the (radio)-oxidation of polyethylene
  in the form of ketones. Starting from the production of alkyl radicals
  (typically by irradiation, reaction 1), the capture of O$_2$ molecules
  follows (reaction 2). The chain reaction can proceed through the formation
  (then decomposition) of hydroperoxides either by unimolecular
reactions 3a-c, or by bimolecular reaction 4 and then unimomolecular or
bimolecular reactions 5-11. The main stable final products are ketones and water
molecules, while hydroperoxides, hydroxyl radicals and peroxy radicals will
take part in further reactions, together with alkyl and alkoxy radicals.}
\label{FigReactionPaths}
\end{figure}

Building on previous works \cite{oluwoye_oxidation_2015,da_cruz_thermo-oxidative_2016} we present in
Figure\ref{FigReactionPaths} a summary of most probable reaction pathways
leading to the oxidation of polyethylene. We assume that alkyl radicals exist
through the C-H bond dissociation by $\gamma$-irradiation (reaction 1). 

\begin{figure}
\includegraphics[width=\columnwidth]{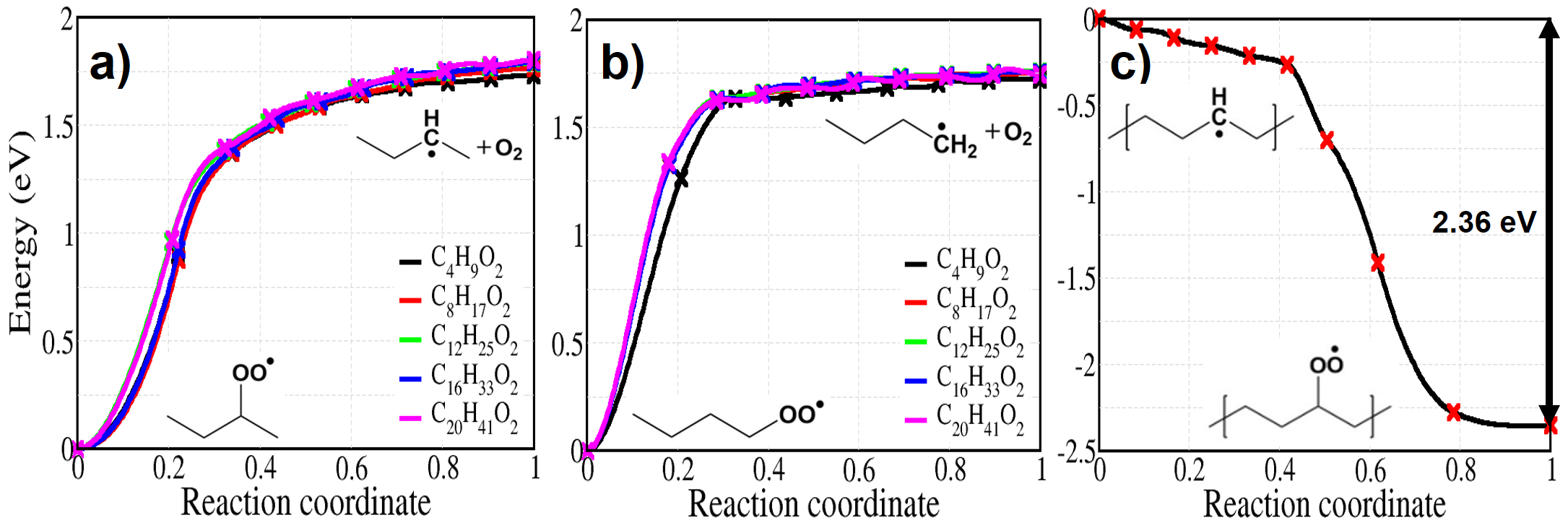}
\caption{The capture of an oxygen molecule by an alkyl radical occurs
  spontaneously, as shown by the energy profiles shown here: a) capture by an
  alkyl radical situated on a internal carbon on alkane molecules of varying
  length b) capture by an
  alkyl radical situated on an end-chain carbon of alkane molecules of varying
  length c) capture by an alkyl radical in a crystalline region of PE.}
\label{FigO2Capture}
\end{figure}

The reported C-H bond energies from methane to propane are in the range of
407.6-439.7 kJ/mol (4.25-4.58~eV) from experiments and 397.5-437.6 kJ/mol (4.14-4.56~eV) from calculations
\cite{kysel_dft_2011,luo_handbook_2004}. After the formation of an alkyl radical, an oxygen molecule
can be captured without any energy barrier at a -$^\bullet$CH- site by forming a peroxy
radical (reaction 2). Figure \ref{FigO2Capture} presents the energy profiles of reaction 2 for both
the molecular and the solid models. 

We checked the reaction for molecules of varying lengths
and both for capture far from chain end (approximately in the middle of the
molecule, Figure~\ref{FigO2Capture}a) or at chain end (Figure~\ref{FigO2Capture}b). Addition of O$_2$ shows barrierless energy profile not only for the various length of alkyl chains but also
for crystalline PE, with high exothermic enthalpies of 1.78~eV (chain centre, average), 1.75~eV (chain end, average), and 2.36~eV (PE crystal), which imply spontaneous $O_2$ capture by an alkyl radical.

\begin{figure}
\includegraphics[width=0.6\columnwidth]{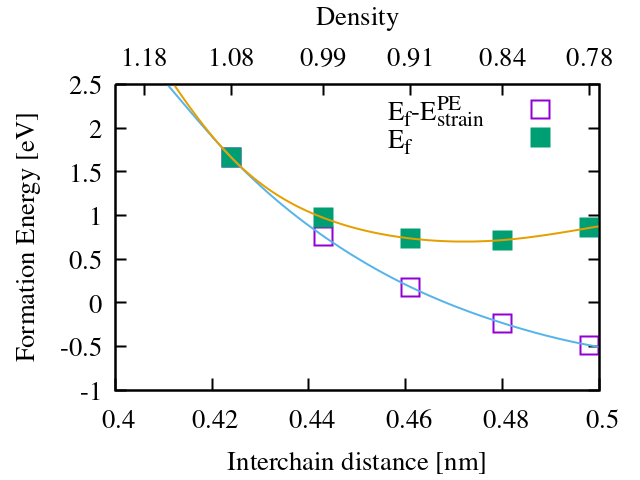}
\caption{Solution energy of an oxygen molecule in crystalline PE as a function
  of the interchain distance. For both the yellow and the blue curves the oxygen molecule is inserted in a model of PE with scaled in plane lattice parameters, so to modify the interchain distance and not the dimensions along the chain. The equilibrium interchain distance is 0.424~nm; the yellow curve gives the standard solution energy (E$_f$), computed with respect to the equilibrium structure; the blue
curve takes as a reference a PE crystal with the same scaled lattice parameters as that of the supercell hosting the molecule, that is we remove the strain energy required to dilate the PE at 0~K (E$_{strain}^{PE}$). The oxygen chemical potential, in both cases, is the energy of an isolated O$_2$ molecule.}
\label{FigO2solution}
\end{figure}

In crystalline PE, at variance with gas phase reactions, the diffusion of
oxygen represents a limiting step. Considering the complex microstructure of
PE, with crystalline and amorphous regions and interfaces between them, a full
account of energy barriers for oxygen diffusion will be a study in
itself. However, useful hints emerge from our calculations of a molecule in
solution into crystalline PE. First, the molecule does not spontaneously
dissociate. Second, the solution energy of the molecule (calculated with
respect to pristine crystalline PE and a gas phase oxygen molecule) is
relatively high at constant volume, but varies considerably when varying the
interchain distance of PE. This can be appreciated from
Figure~\ref{FigO2solution}, which suggest a behaviour similar to O$_2$ in
amorphous silica \cite{bongiorno_oxygen_2002}: diffusion takes place in the
amorphous more easily than in the crystal, thanks to a wide distribution of
interchain spacings; this view corroborates the usual assumption that oxygen
in PE essentially diffuses through the amorphous regions\cite{michaels_flow_1961}.

The
typical activation energies associated to oxygen dissolving and diffusing for solution and diffusion in PE range from 0.35 to
0.45 eV \cite{michaels_flow_1961}. However, in this case, this activation energy
might be actually related to permeation instead of diffusion, because, at
variance with SiO$_2$ where bottleneck exists between the voids where the molecule
can easily sit, in PE the channels between alkyl chains constitute an easy
diffusion path. To check this we performed a NEB calculation for the migration
of an oxygen molecule between two equivalent neighbouring insertion sites along
the channels and found a migration barrier on the order of 0.1~eV. To corroborate this results we also performed an ab initio molecular dynamics simulation which, although not long enough to provide an accurate value of the free energy barrier, featured a number of jumps of the molecule consistent with the migration barrier calculated with the NEB method.
This result, combined with those in Figure~\ref{FigO2solution}) suggest that the previously cited permeation activation energy (0.4~eV~\cite{wang_ethylene_1998}) stems essentially from thermodynamics (i.e., the solution energy) and not from kinetic barriers, and probably represents an average of the solution energy over the available insertion sites, mainly in the amorphous and in the interface regions. In real samples we can imagine a variety of reasons why the concentration of O$_2$ molecules, in both the amorphous and the crystalline regions, is not the equilibrium one: quenching from higher temperatures, recrystallization during irradiation, extrusion or material synthesis under higher oxygen pressure.
For these reasons, the presence of O$_2$ cannot be categorically excluded in the crystalline regions.

\subsection{Formation of hydroperoxides}
\label{FormHPOx}

After the capture of O$_2$ by an alkyl radical (reaction 2), the reaction
pathway branches into several possible channels towards the formation of
hydroperoxide groups. Let us first consider the formation of a hydroperoxide
by H-abstraction from the same alkyl chain; depending on its positon it is
labeled $\gamma$, $\beta$, or $\alpha$ (reaction 3a, 3b, and 3c). Reaction 3c
does not form the hydroperoxide but a ketone and a water molecule, because its
intermediate configuration, which is $\alpha$-alkyl-hydroperoxy radical,
dissociates in less than 20 $\mu$s \cite{vaghjiani_kinetics_1989,gugumus_thermolysis_2000}.
We do not
consider H abstraction from more distant sites (e.g., from $\delta$-position)
because the alkyl chain is hardly bent in crystalline PE, which would
presumably lead to unfavourable pathways. 
\begin{figure}
\includegraphics[width=\columnwidth]{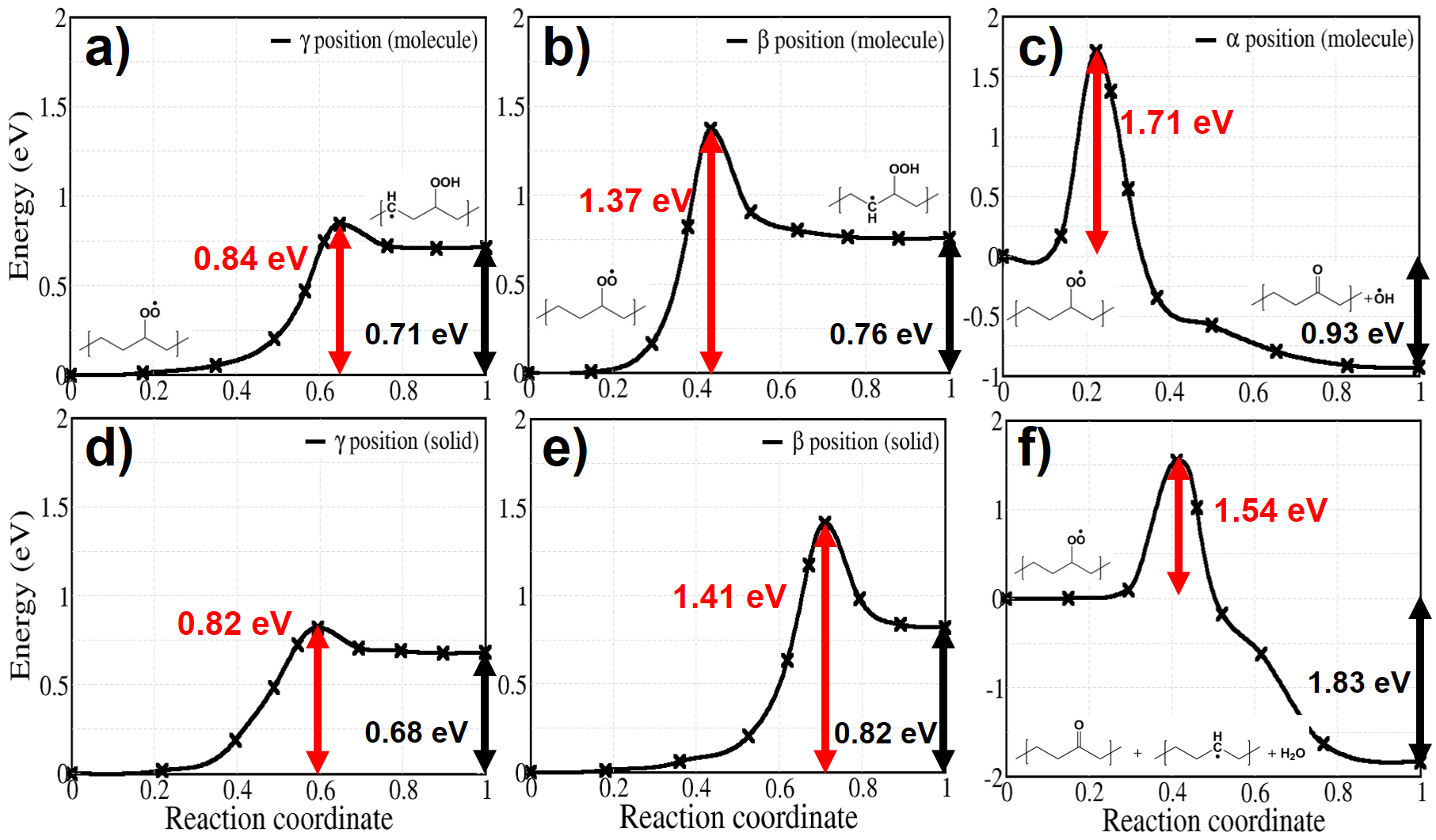}
\caption{Energy profiles for reactions labeled 3a-c in
  Figure~\ref{FigReactionPaths}. Panels a-c: H-abstractions from
  $\gamma$,$\beta$,$\alpha$, respectively, in a molecular model. Panels d-f:
  analogous reactions for crystalline PE.}
\label{FigReactions3}
\end{figure}

\begin{figure}
\includegraphics[width=\columnwidth]{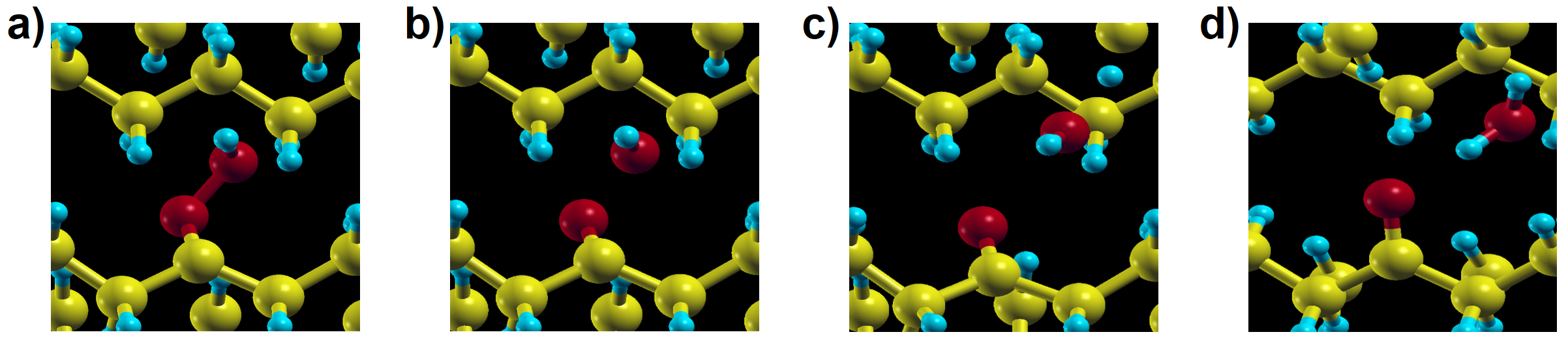}
\caption{Representative structures of the reaction whose energy profile is
  shown in panel f of Figure~\ref{FigReactions3}, corresponding to reaction 3c
  of Figure~\ref{FigReactionPaths} as it occurs in crystalline PE.}
\label{FigSolidR3snapshots}
\end{figure}

Figure~\ref{FigReactions3} shows
calculated activation energies for both molecular and solid models. The
activation energy of H abstraction from the $\gamma$ position has the lowest
values among the H abstractions: it amounts to 0.84~eV and 0.82~eV for
the molecular and the solid model, respectively (Figs.~\ref{FigReactions3}a and
\ref{FigReactions3}d). As a hint on the influence of density on the energy barriers in solid PE, we calculated the barrier of reaction 3a at several interchain distances, from 0.42 to 0.48 nm; the variation of the energy barrier does not exceed 0.05 eV. 

H abstraction from $\beta$ position shows higher energy
barriers of 1.37~eV and 1.41~eV for the molecular and the solid model, respectively
(Figure \ref{FigReactions3}b and \ref{FigReactions3}e). $\gamma$ and $\beta$
hydrogen abstraction both are endothermic, which means that a reverse reaction
is more likely than the forward one, reducing the effective rate constant of hydroperoxide production. In contrast, $\alpha$ hydrogen abstraction is exothermic, with the
final equilibrium structure having much more stable energy than for the other
two mentioned reactions. Although this reaction would directly lead to ketone
products as observed in thermo- or photo-oxidation of PE, it has a high
activation energy (1.71~eV for the molecular model and 1.54~eV for the
crystal) and is thus unlikely to occur in normal conditions. In contrast with
reaction 3a and 3b, showing similar behaviour in the solid and for the gas phase
molecules, reaction 3c in crystalline PE leads to a different final product:
during the reaction a hydroxyl radical is produced from P$^\bullet$O-OH dissociation
and reacts with another neighbouring polymer chain, forming an alkyl radical
and a water molecule in crystalline PE. This additional process occurs
spontaneously, as seen from the structural relaxation in
Figure~\ref{FigSolidR3snapshots}. Although the instability of
$\alpha$-alkyl-hydroperoxy radical and the reactivity of hydroxyl radical have
already been investigated\cite{vaghjiani_kinetics_1989,gugumus_thermolysis_2000,vereecken_computational_2004} in simple
molecular systems, not much has been done concerning these reactions in
crystalline polymers nor their role in a global kinetic pathway. Regarding the
hydroxyl radical, its role in catalysing hydroperoxide formation will be
discussed in detail with reaction~9. Reactions involving additional steps are marked by rectangles in Figure~\ref{FigReactionPaths}.
\begin{figure}
\includegraphics[width=\columnwidth]{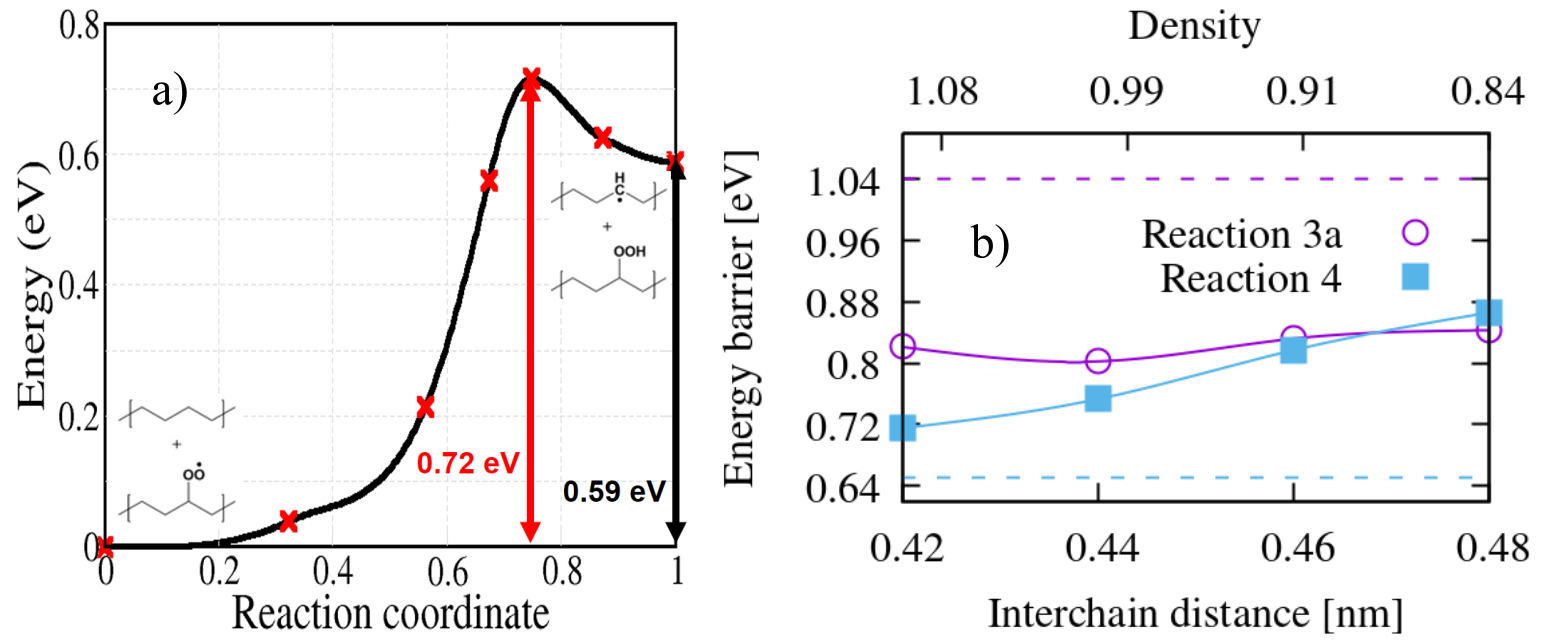}
\caption{a) The energy profile calculated for the bimolecular reaction~4 of
  Figure~\ref{FigReactionPaths} as calculated in crystalline PE. b) The influence of the PE interchain distance on the energy barriers of bimolecular reactions 3a and 4. The dashed lines are the energy barriers calculated for the same reactions in the lamellar interface structure.}
\label{FigR4}
\end{figure}

 On the other hand, hydroperoxides can be formed also when the peroxy radical grabs
 an H atom from an adjacent polymer chain in the system. In order to simulate
 this intermolecular abstraction we chose the H atom which is closest to the
 peroxy radical, in crystalline PE (reaction~4);  for all bimolecular
 reactions we limited our study to the crystalline system. The calculated
 activation energy is 0.72~eV (Figure \ref{FigR4}). This reaction is competitive with $\gamma$-intramolecular H abstraction (Figure\ref{FigReactions3}a,d) because they both contribute to the formation of a hydroperoxide. In previous kinetic modelling this sort of reaction was treated with an activation energy of 0.76~eV\cite{khelidj_oxidation_2006,khelidj_oxidation_2006-1}, not far from our calculated value.

 The energy barrier of intermolecular reaction 4 depends more strongly on the interchain distance than the intramolecular reaction 3a: the value of 0.72~eV corresponds to the zero temperature theoretical equilibrium structure (interchain distance 0.42~nm), while if we stretch the interchain distance to the value of 0.46~nm (corresponding approximately to the room temperature density of LDPE) the energy barrier is raised to 0.82~eV. This gives a hint on the distribution of energy barrier for hydroperoxide formation by intermolecular H-abstraction in the amorphous region, where interchain distance can locally vary.
 Apart from density, however, the local conformation of the chains can also play a role; for the sake of illustration we computed the energy barrier of reactions 3a and 4 using the lamellar model shown in Figure~\ref{FigLamellar}. The resulting energy barriers (see dashed lines on Figure~\ref{FigR4}b) seems to point to the fact that bent chains substantially raise the barrier for the unimolecular reaction 3a, while lower slightly the energy barrier of the bimolecular reaction~4; of course this will somewhat depend on the choice of the interface site where the peroxy radical is grafted, in our case the one shown in Figure~\ref{FigLamellar}. 


\subsection{Decomposition of hydroperoxides}
\label{DecompHPOx}

\begin{figure}
\includegraphics[width=0.6\columnwidth]{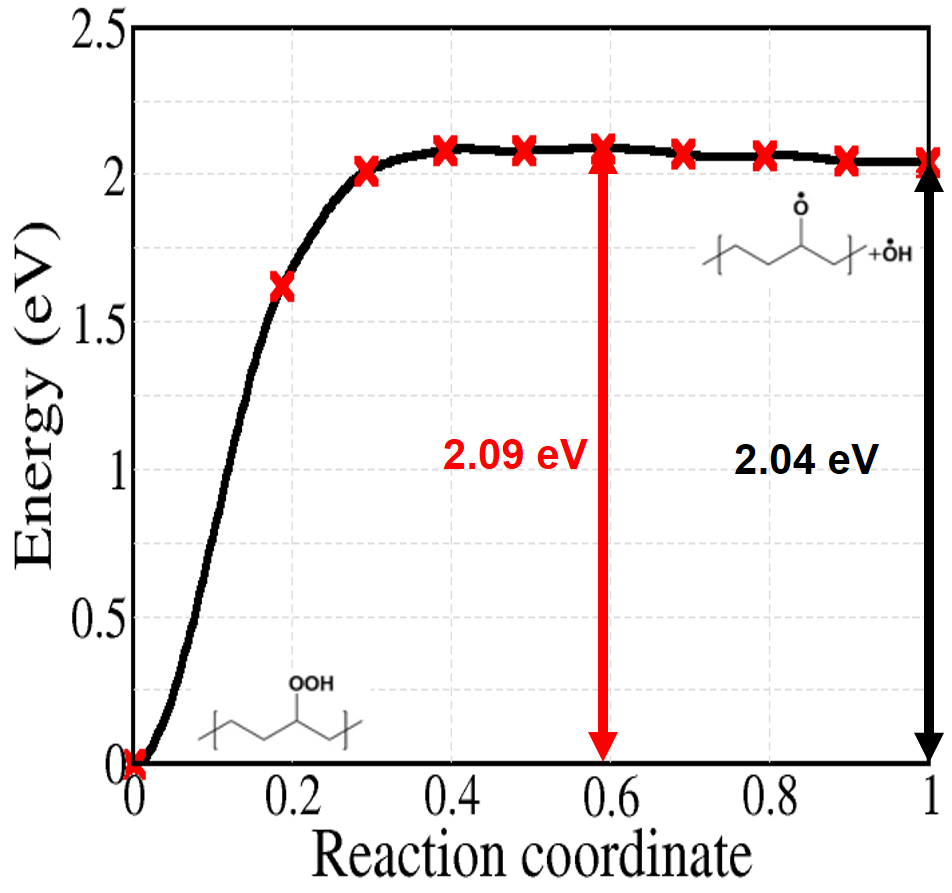}
\caption{Calculated energy profile for reaction 5 of Figure~\ref{FigReactionPaths}.}
\label{FigR5}
\end{figure}

\begin{figure}
\includegraphics[width=0.6\columnwidth]{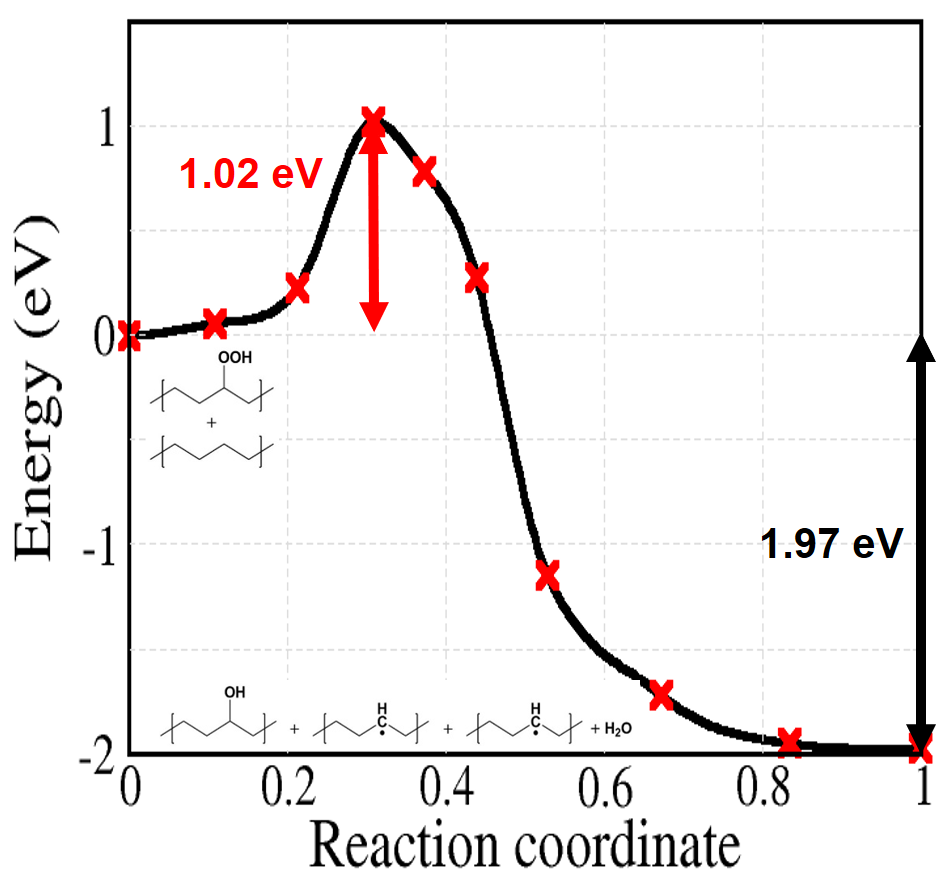}
\caption{Calculated energy profile for reaction 6 of
  Figure~\ref{FigReactionPaths} in a model of crystalline PE. The calculated outcome is here different from the
  commonly accepted one, giving instead two alkyl radicals and an alcohol, plus
a water molecule.}
\label{FigR6}
\end{figure}

\begin{figure}
\includegraphics[width=\columnwidth]{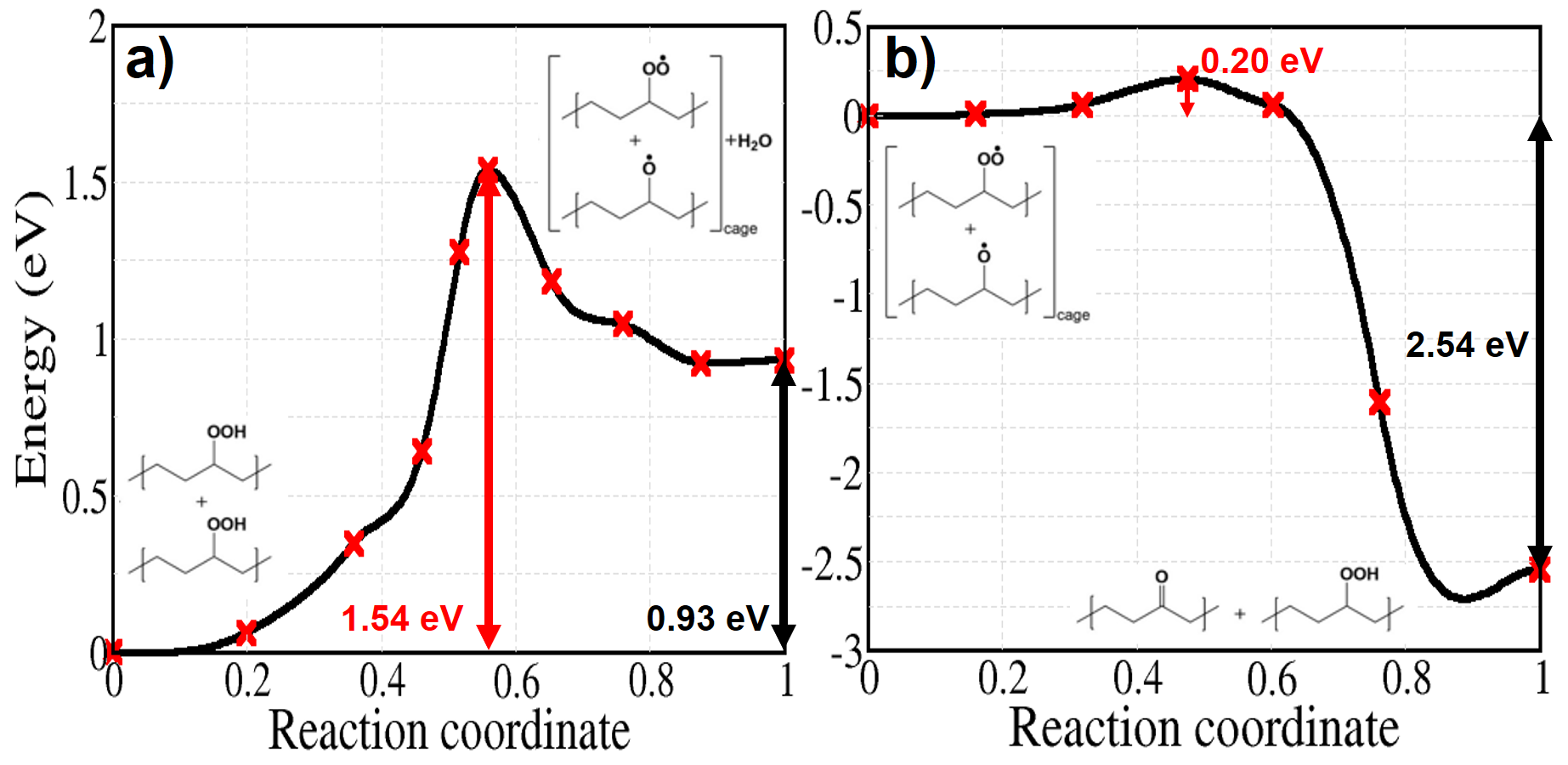}
\caption{Energy profiles for reactions 7 (panel a) and 8 (panel b) of
  Figure~\ref{FigReactionPaths} calculated for crystalline PE.}
\label{FigR7-8}
\end{figure}

Reaction~5 to 9 of Figure~\ref{FigReactionPaths} are related to the decomposition
of hydroperoxides, which can occur both by unimolecular or bimolecular
processes. At first, we can simply consider the unimolecular PO-OH bond
thermal dissociation (reaction~5). As shown in Figure~\ref{FigR5}, this
reaction shows the highest activation energy (2.09~eV) among the reactions in
Figure~\ref{FigReactionPaths}. From the experimental kinetics of hydroperoxide
decomposition, the contribution of this reaction is only up to 2\% for
temperatures below 200$^o$C, indicating the importance of pseudo-unimolecular
POOH decomposition such as reaction~6 when hydroperoxide concentration is low ([PH] $>>$ [POOH]) \cite{de_sainte_claire_degradation_2009,gugumus_thermolysis_2000,gugumus_re-examination_2002}. 
 When the concentration of POOH is small enough, reaction~6 (Figure\ref{FigR6}) would also prevail
 over reaction~7 (Figure\ref{FigR7-8}a).  However, for the former reaction, the traditionally accepted
 reaction pathway shown in Figure~\ref{FigReactionPaths} comes into question. In fact, according to
 our calculations, it proceeds to a different
 product: the alkoxy radical in crystalline PE
 spontaneously abstracts an H atom from another alkyl chain, forming an alcohol as
 shown in Figure~\ref{FigR6} (this further, unexpected, step is shown in the box after reaction 6 in Figure~\ref{FigReactionPaths}).
 
The obtained activation energy is 1.02~eV. This reaction, among the considered
ones, is the one that forms the most alkyl radical chains with comparatively
low activation energy. These alkyl radicals can diffuse along or across the
chain. Otherwise, with the presence of oxygen, alkyl radicals do not propagate
and grab oxygen as reaction 2 by forming peroxy radical defects.

\begin{figure}
\includegraphics[width=\columnwidth]{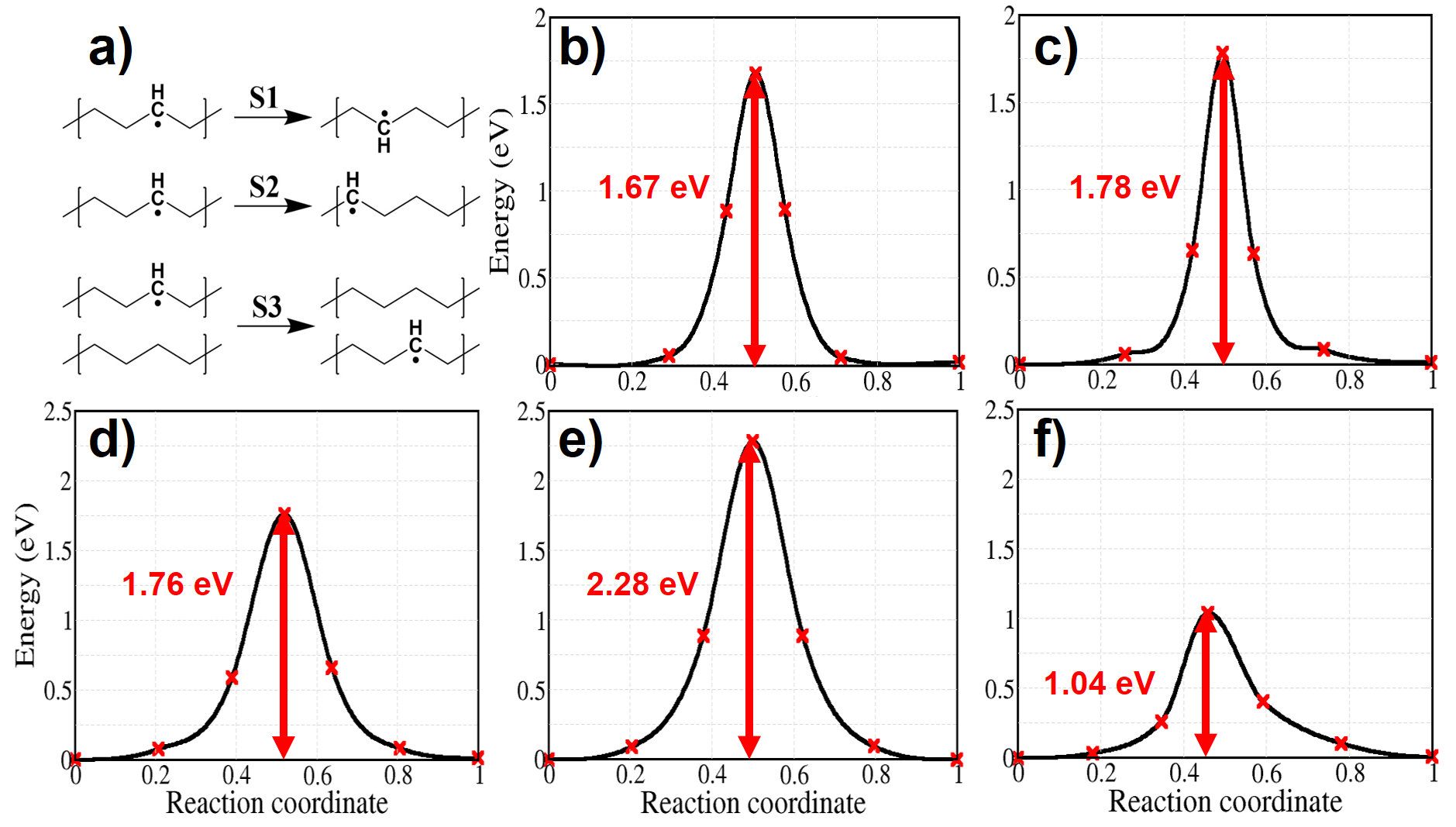}
\caption{Energy profiles for the migration of an alkyl radical. Panel a):
  schemes for nearest neighbor (S1) and second nearest neighbor (S2) migration
along the alkane chain, and from a chain to a neighboring one (or across the
chains, S3). Panels b,c: energy profiles for migrations S1-S2 for the molecular model. Panels d,e:
migrations along the chain for the solid model; panel f: migration across the
chains (S3) in crystalline PE. }
\label{FigAlkylMigra}
\end{figure}

We calculated the
activation energy for the migration of an alkyl radical through three possible atomic jumps and we show it in
Figure~\ref{FigAlkylMigra}. Shimada {\it et al.}\cite{shimada_free_1977,shimada_relation_1981} compared the
decay rate of alkyl radicals trapped in the urea-polyethylene complex and in
solution grown crystals. Although the mobility of  PE chains is higher in the
urea-polyethylene complex, alkyl radicals decay at slower rate here than in the
solution grown crystals, where interchain motion is reduced. This finding
implies that the rate of alkyl radical propagation {\sl across} the chains is much
faster than that along the chain in polyethylene crystals. In agreement with
such a result we find an activation energy for alkyl radical migration from
chain to chain of 1.04 eV, which is much lower than the activation energy for
alkyl migration along the chain both in the molecular and the solid models
(see Figure~\ref{FigAlkylMigra}).
 The bimolecular decomposition of POOH (reaction 7) is relevant when the
 concentration of hydroperoxides is  sufficiently high or when an
 inhomogeneous distribution of hydroperoxides leads to local accumulations of
 these groups and may become  the main  free radical formation
 reaction\cite{zolotova_mechanism_1971}. Figure~\ref{FigR7-8}a shows the energy profile of
 reaction 7, which has a relatively high energy barrier of 1.54 eV. After the
 hydroperoxide decomposition the reaction takes place depending on the
 existence of the so called cage effect ---i.e., the proximity of the two
 radicals--- and their tendency to diffuse away from each other along the
 alkyl chains. When the radicals are trapped in the cage
 without diffusing, the reaction forms a hydroperoxide and a ketone by
 abstracting the tertiary H atom of the alkoxy radical. While this process has
 been supposed to  happen  with no activation energy, our calculation shows a
 small activation energy of 0.20 eV (Figure~\ref{FigR7-8}b)
 \cite{gugumus_physico-chemical_2005,nangia_kinetics_1980}. 

 Apart from the previous reaction, a hydrogen from a secondary hydroperoxide can be attacked by other
 radical species resulting from the products
 of other PE oxidation reactions. We considered three types of radicals: i) a hydroxyl ii) a peroxy radical, and iii) an alkyl radical.
 The hydroxyl radical has a high reactivity and easily abstracts a H atom
 from the alkane. De Saint-Claire\cite{de_sainte_claire_degradation_2009} reported a rate constant
 (10$^{12-13}$ cm$^3$ mol$^{-1}$ s$^{-1}$) of H atom abstraction by hydroxyl
 radical, which is much higher than that of other reactions, but coherent with
 previous experimental estimations for $^\bullet$OH radical reactions with alkanes\cite{hickel_absorption_1975,rudakov_low_1981}. The induced
 decomposition by the radicals amounts up to 54\% of the overall hydroperoxide
 decomposition channel in thermo-oxidative conditions. Reactions with hydroxyl
 radical in crystalline PE are essential because the radical can exist not
 only from products of reactions but also from outside of crystalline PE,
 namely the amorphous phase. Moreover, the permeability of PE to
 the hydroxyl radical is expected to be comparatively higher than to oxygen~\cite{Rogers-Comyn}, because
 the $^\bullet$OH radical size is smaller than O$_2$. Therefore, we tested these reactions by
 putting a hydroxyl radical next to a hydroperoxide on primary and tertiary
 sites for reaction 9a and 9b, respectively. Reaction 9a produces a peroxy
 radical and a water molecule. For reaction 9b, firstly, the hydroxyl radical
 abstracts the tertiary H atom of the hydroperoxide. As an intermediate state an
 $\alpha$-alkyl-hydroperoxy radical and a water molecule are formed, then the
 P$^\bullet$O-OH bond is decomposed immediately by forming a ketone and a hydroxyl
 radical. Both these reactions occur spontaneously. 
In order to check our initial position we performed relaxations constraining the distance between the tertiary H atom of the secondary hydroperoxide and the oxygen atom of the hydroxyl radical at various intermediate values. As for the oxygen capture from an alkyl radical, our calculations show that no barrier needs to be overcome to trigger this reaction.
 
\begin{figure}
\includegraphics[width=\columnwidth]{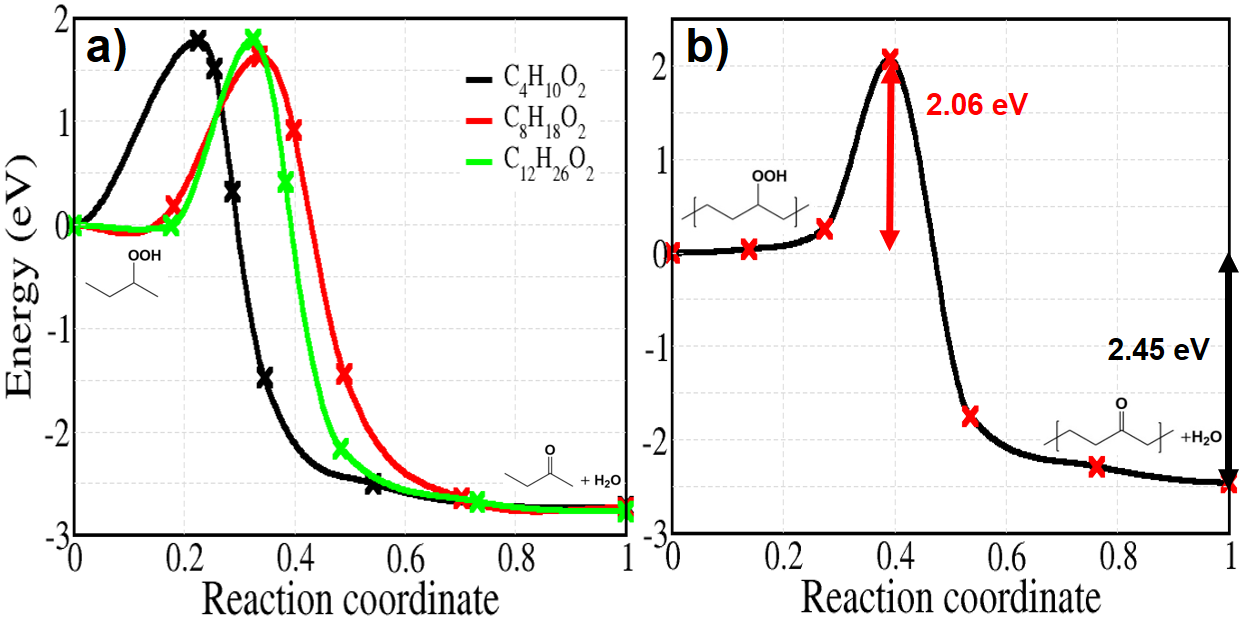}
\caption{Energy profiles calculated for reaction 9c of
  Figure~\ref{FigReactionPaths}. Panel a): molecular models of varying sizes. Panel
b): crystalline PE.}
\label{FigR9c}
\end{figure}

We also verified by hybrid functional
calculations the energy profile of those relaxations, confirming the
spontaneous H-abstraction by the $^\bullet$OH radical. 
\begin{figure}
\includegraphics[width=\columnwidth]{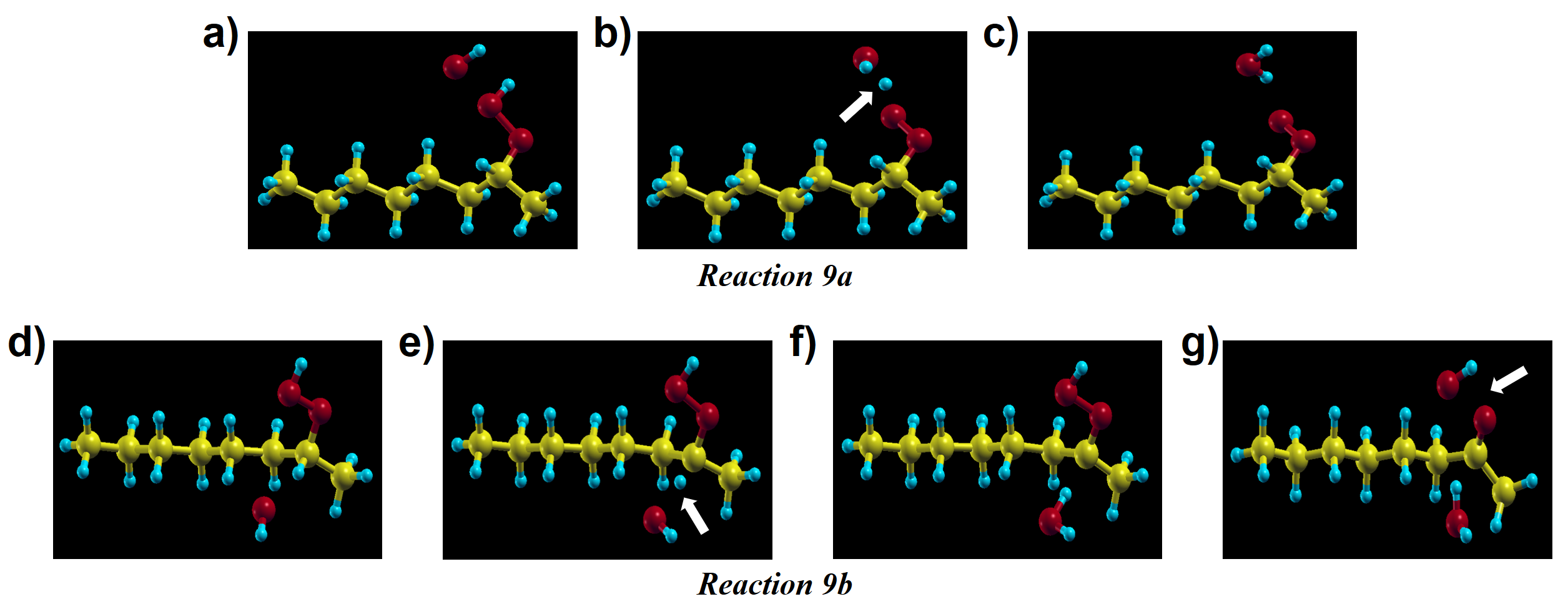}
\caption{Intermediate steps of reactions 9a (panels a-c) and 9b (panels d-g) of
  Figure~\ref{FigReactionPaths}, showing two possible channels of hydroperoxide decomposition due to
  the action of a hydroxyl radical in a molecular model. The different outcome of the two reactions
stems from the initial position of the hydroxyl relative to the hydroperoxide group.}
\label{FigR9steps}
\end{figure}
Figure \ref{FigR9steps}a-c and \ref{FigR9steps}d-g show intermediate steps
along the structural relaxations corresponding to reactions 9a and 9b,
respectively. A hydroxyl radical abstracts a H atom spontaneously, without any
barrier.  In
particular, we could see that a new hydroxyl radical is again present in the
final state of reaction 9b; this $^\bullet$OH radical can react with other molecules
or alkyl chains successively. This is important because, as we confirmed
dealing with reaction 3c in crystalline PE, hydroxyl radical can be available
for these reactions to proceed spontaneously. Intermediate steps of the
structural relaxation corresponding to reaction 9a and 9b in crystalline PE
are presented in Figure~\ref{FigR9solidsteps}. While the reaction 9a in
Figure~\ref{FigR9solidsteps}a-c shows the
same products as its molecular model in Figure~\ref{FigR9steps}a-c, the decomposed hydroxyl
radical from the hydroperoxide in reaction 9b abstracts a H atom from another alkyl
chain producing an alkyl radical chain (Figure~\ref{FigR9solidsteps}d-i).  The overall reaction proceeds as follows:
\begin{equation}
POOH+ {}^\bullet OH\rightarrow P^\bullet OOH+H_2O\rightarrow P=O+P^\bullet+2H_2O .
\label{Eq9bsolid}
\end{equation}
Note that reactions 9a and 9b, as well as 3c, could occur also in amorphous regions, similarly to the crystal, whenever two polymer chains are close enough.
\begin{figure}
\includegraphics[width=\columnwidth]{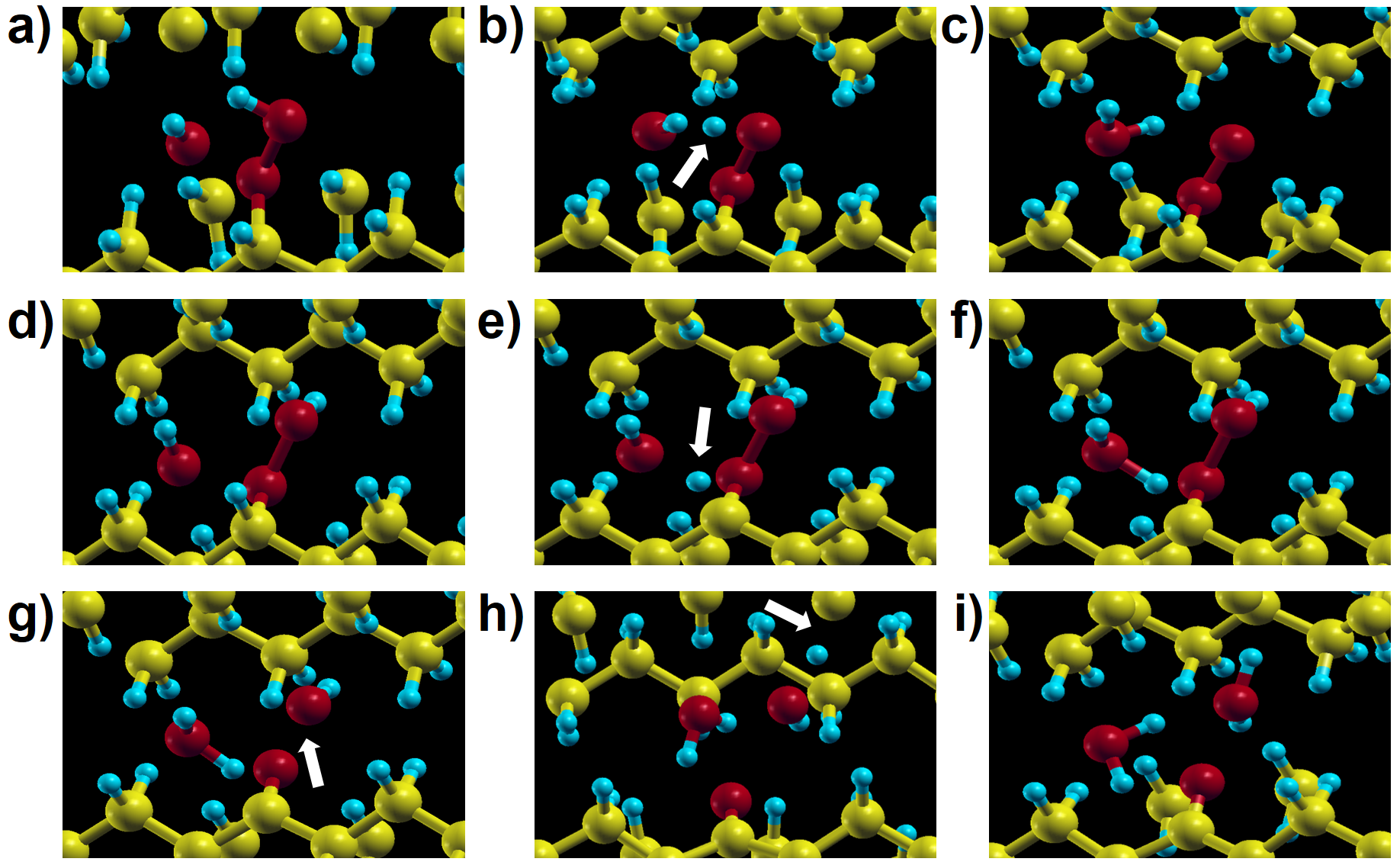}
\caption{Intermediate steps of reactions 9a (panels a-c) and 9b (panels d-i) in
  Figure~\ref{FigReactionPaths} in crystalline PE. The outcome of reaction 9b in
  the solid is different from the analogous one in the molecular model
  (Figure~\ref{FigR9steps}d-g), because here the remaining $^\bullet$OH radical reacts
with a neighbouring chain (see the arrow in panel h) giving one more water
molecule (panel i).}
\label{FigR9solidsteps}
\end{figure}
 Finally, we deal with reaction 9c of Figure~\ref{FigReactionPaths}, starting from
 a hydroperoxide without hydroxyl radical and giving a ketone and water in the
 final state; the corresponding activation energy is calculated by varying the
 number of carbon from 4 to 12 for comparison (Figure~\ref{FigR9c}a). In
 contrast with the previous reactions 9a and 9b of spontaneous H abstraction,
 the barriers here are quite large, regardless the size of the molecules,
 showing an average value of 1.73~eV.  The activation energy for the solid
 model also has the high value of 2.06 eV (Figure~\ref{FigR9c}b). The energy is
 not far from that of reaction~5, describing the dissociation of a PO-OH bond.  

\begin{figure}
  \includegraphics[width=\columnwidth]{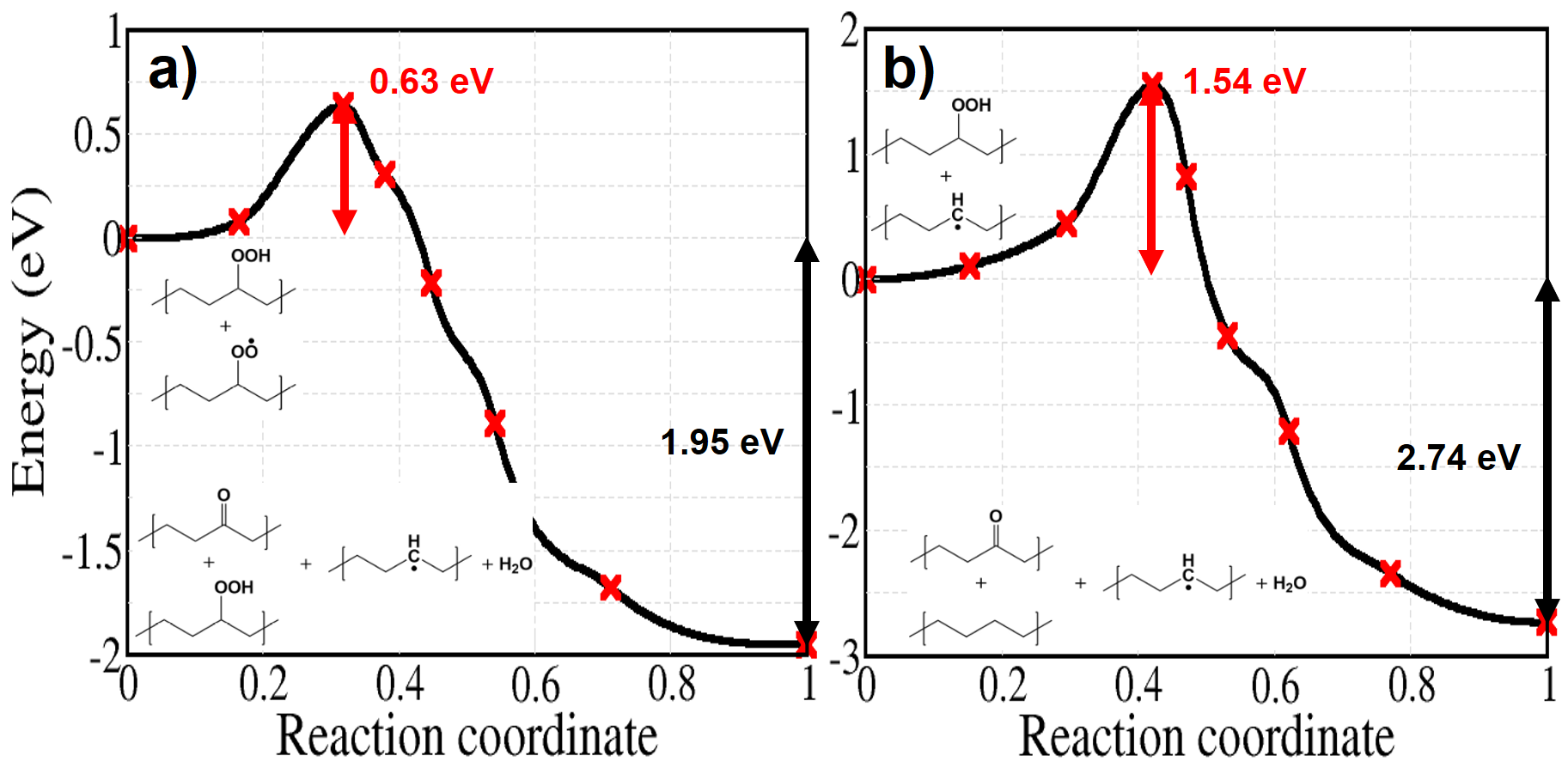}
  \caption{Energy profiles for hydroperoxide decomposition reactions induced by a) a peroxy radical b) an alkyl radical, sitting on a nearby polymer chain.}
\label{FigOtherRadicalReactions}
\end{figure}

Energy profiles of free radical induced decomposition of hydroperoxides are shown in Figure\ref{FigOtherRadicalReactions}. They both attack a tertiary hydrogen atom of a secondary hydroperoxide in the same manner as in reaction 9b. Although the activation energy of reaction~10 is higher than that for hydrogen abstraction by hydroxyl radicals, we can observe that a hydroperoxide is regenerated during the reaction; this is important during the initial stages of the oxidation\cite{petruj_mechanism_1980,gugumus_physico-chemical_2005-1}, when the formation of ketones proceeds at a constant rate. In contrast with reaction 9b and 10, the activation energy of reaction 11 is high and amounts to 1.54~eV. This implies that hydrogen abstraction at a tertiary site is not seriously affected by an alkyl radical sitting on a nearby polymer chain.

If we consider the decomposition of hydroperoxides in absence of 
radical attacks, the easiest path is through reaction~6, whose activation energy is
1.02~eV, all others having higher activation energies. In contrast, a hydroxyl
radical spontaneously removes the H atom of hydroperoxides, as well as other H
atoms in alkyl chains. 
In other words, all H abstractions related to hydroxyl
radical (reaction 3c, 9a, and 9b) showed barrierless energy
profiles. Especially, during the reaction 9b, a new hydroxyl radical is again
generated after the abstraction of tertiary hydrogen, potentially triggering a
chain reaction with other alkyl radicals. Besides, a hydrogen abstraction by a peroxy radical also involves an energy barrier which is low compared to other hydroperoxide decomposition reactions, by forming an additional hydroperoxide as a product. Therefore, the role of peroxy and, even much more, hydroxyl
radicals, causing chain reactions, cannot be overlooked for the formation of ketones or other radical species and seems to be
crucial for the oxidative degradation PE.   

\section{Discussion}
\label{Discu}

As a basis for discussion all calculated activation energies are tabulated in
Table~\ref{TabActivationEnergies}. 

\begin{table}
\caption{Activation energies calculated in the present paper for the reactions
  shown in Figure~\ref{FigReactionPaths}. Energies are in eV.}
\begin{tabular}{c c }
Reaction & Activation energy [eV] \\ \hline
\begin{tabular}{l c}
Description & label in Figure~\ref{FigReactionPaths} \\ \hline
oxygen capture & 2\\ 
$\gamma$-H abstraction & 3a\\
$\beta$-H abstraction & 3b \\
$\alpha$-H abstraction  & 3c \\
 bimolecular H-abstraction & 4\\
 unimolecular PO-OH bond cleavage  & 5 \\
 pseudo-unimolec. POOH decomposition & 6\\
 bimolecular POOH disproportionation & 7\\
 bimolecular alkoxy/peroxy reaction & 8\\
POOH decomposition by $^\bullet$OH (1) & 9a\\
POOH decomposition by $^\bullet$OH (2) & 9b\\
unimolecular POOH decomposition & 9c\\
POOH decomposition by POO$^\bullet$ & 10 \\
POOH decomposition by P$^\bullet$ & 11\\
\end{tabular}
& 
\begin{tabular}{c c}
molecule & crystal \\ \hline
no barrier & no barrier \\
0.84 & 0.82 \\
1.37 & 1.41 \\
1.71 & 1.54 \\
 -- & 0.72 \\
2.09 & -- \\
-- & 1.02 \\
-- & 1.54 \\
-- & 0.2 \\
no barrier & no barrier \\
no barrier & no barrier \\
1.73 (average) & 2.06 \\
-- & 0.63 \\
-- & 1.54 \\
\end{tabular} \\ \hline
\end{tabular}

\label{TabActivationEnergies}
\end{table}

Assuming the production of alkyl radicals, and the spontaneous capture of
oxygen molecules by the latter, as shown in section~\ref{OCapture}, the next
important step is the formation of hydroperoxides. The reactions showing the
lowest barriers for this process are reactions 3a and 4. However, the reverse
barrier is small (only a few tenths of an eV), lower than the forward barrier
and of the diffusion barrier for the alkyl radical, suggesting that the effective
rate of these reactions is low, because the hydroperoxide can easily decompose
in its constituents, peroxy radical and a restored alkyl chain. 
Another possible direct channel of ketone formation, without hydroperoxide
intermediates, could possibly stem
directly from reaction 8 of Figure~\ref{FigReactionPaths}, provided that alkoxy
radicals are available and get close to peroxy ones. This might be
facilitated by the reaction of hydrogen molecules with peroxy radicals

Apart from the reverse reactions of 3a and 4, hydroperoxides can decompose
following reaction 6 (barrier 1.02), possibly followed by an oxygen
capture by the alkyl and a subsequent reaction 8 (barrier 0.2~eV). The highest barriers involved in those processes are
not very easy to overcome at room temperature, although for reactions 3a and 4 the
rate might be enhanced by a sufficiently large concentration of alkyl
radicals.
Hydroperoxide decomposition might be enhanced by the presence of hydroxyl
radicals, however their production through reactions 3c and 5 is hindered by a
high energy barrier ($>$ 1.7~eV).

The relative importance of the various formation and decomposition reactions of hydroperoxides depends thus on their initial concentration and that of various types of radicals. A direct experimental probe of the concentration of the various species, including in particular hydroperoxides, would be of great help to further unravel the degradation mechanisms in polyethylene.

\section{Summary and Conclusions}
\label{Conclu}

Degradation mechanisms of oxidation PE were investigated by calculating
reaction barriers in small molecules and crystalline polyethylene. We divided
our study in three main parts: i) oxygen capture by an alkyl radical, ii)
formation of hydroperoxides, and iii) decomposition of hydroperoxides. First,
oxygen reacts with alkyl radicals spontaneously, even though diffusion through
crystalline PE might constitute a limiting step.

After the formation of peroxy radicals, hydroperoxides are formed by intramolecular
or intermolecular H abstractions. We considered the H abstraction from
$\alpha$, $\beta$, and $\gamma$ positions. Among the intramolecular reactions,
abstraction from the $\gamma$ position has the lowest activation energies, which are 0.84 eV and 0.82 eV for molecular and solid model, respectively. The energy barrier of intermolecular H abstraction is 0.72~eV. Therefore, competition between these two reaction pathways is probable in crystalline PE.

Although, as we show, the solubility of oxygen in crystalline regions is expected to be much lower than in the amorphous, for many bimolecular reactions the energy  barriers in the crystal should provide a reasonable estimate for the amorphous regions, as we have shown by varying the interchain distance.

 In contrast with the typical assumptions made in the literature, our
 calculations show that some reactions such as reaction 3c ($\alpha$-H
 abstraction), 5-6 (PO-OH bond cleavage), and 9b (POOH decomposition
 through $^\bullet$OH radical), have
 different outcomes whether the reaction occurs in crystalline (and probably also amorphous) PE or in an
 isolated alkane molecule. In particular, in presence of hydroxyl radical,
 further reactions take place and the activation energy of decomposition of
 hydroperoxide varies largely. Without the radical, the reaction of
 bimolecular POOH decomposition leading to an alcohol, an alkyl radical and a
 water molecule  has the minimum activation energy of 1.02 eV. For isolated
 molecules, the activation energy of the unimolecular dissociation of PO-OH is
 much higher, 2.09~eV. In contrast, H abstraction by hydroxyl radical is
 spontaneous regardless of its position, both in the crystal and on an alkane molecule. Especially, H abstraction
 from a hydroperoxide in tertiary site leads to successive reactions with other alkyl
 chains and finally leaves an alkyl radical chain. Comparing the energy of C-H
 bond dissociation in pure crystalline PE, which is up to 439.7 kJ/mol, this
 mechanism forming alkyl radical chain is much more favourable. Therefore, we
 conclude that, even in presence of small concentrations of hydroxyl radicals, the reactions they induce may lead to a critical degradation of crystalline PE. 

We believe that our study of
activation energies and reaction processes provides information at the atomic
scale which can be hardly obtained by experiments, and  gives insights on
polymer degradation mechanism which are useful for further kinetic studies.

\vspace{6pt} 



\authorcontributions{Conceptualization, Y.A., G.R., and X.C.; methodology, G.R. and Y.A.; software, G.R. and Y.A.; validation, G.R., Y.A., and X.C.;  formal analysis, Y.A. and G.R.; investigation, Y.A.; resources, G.R.; data curation, Y.A.; writing---original draft preparation, Y.A.; writing---review and editing, G.R., Y.A., and X.C.; visualization, Y.A. and G.R.; supervision, G.R.; project administration, G.R.; funding acquisition, G.R. and X.C. All authors have read and agreed to the published version of the manuscript.}

\funding{This publication was prepared in the context of the TeaM Cables project. This project has received funding from the Euratom research and training programme 2014-2018 under grant agreement No 755183.
This work was granted access to the HPC resources
of TGCC and IDRIS under the allocation 2020A0090906018 made by GENCI and under the allocation by CEA-DEN.
}

\institutionalreview{Not applicable}

\informedconsent{Not applicable}

\dataavailability{The data presented in this study are available on request from the corresponding author.} 

\acknowledgments{Muriel Ferry is gratefully aknowledged for a critical reading of the manuscript.}

\conflictsofinterest{The authors declare no conflict of interest.} 

\end{paracol}
\reftitle{References}


\externalbibliography{yes}
\bibliography{PolyDBZ_UTF8}


%


\end{document}